\documentclass{ws-ijmpe}

\begin{document}

\markboth{Tetsuo Hyodo}{Chiral dynamics, structure of $\Lambda(1405)$, 
and $\bar{K}N$ phenomenology}

\catchline{}{}{}{}{}

\title{CHIRAL DYNAMICS, STRUCTURE OF $\Lambda(1405)$, \\
AND $\bar{K}N$ PHENOMENOLOGY}

\author{\footnotesize TETSUO HYODO}

\address{Physik-Department, Technische Universit\"at M\"unchen, 
D-85747 Garching, Germany, and\\
Yukawa Institute for Theoretical Physics, Kyoto University, 
Kyoto 606--8502, Japan \\
thyodo@ph.tum.de}

\maketitle

\begin{history}
\received{(received date)}
\revised{(revised date)}
\end{history}

\begin{abstract}
  We investigate the structure of the $\Lambda(1405)$ resonance and 
  $\bar{K}N$ phenomenology in the perspective of chiral SU(3) dynamics. 
  Utilizing the chiral coupled-channel approach which well describes the 
  $\bar{K}N$ scattering observable, we perform three different analyses to 
  clarify the structure of the $\Lambda(1405)$ resonance. The results 
  consistently indicate the meson-baryon molecule picture of the 
  $\Lambda(1405)$. We argue the consequence of the chiral dynamics in 
  $\bar{K}N$ phenomenology and the antikaon bound state in nucleus, 
  emphasizing the important role of the strong $\pi\Sigma$ interaction.
\end{abstract}

\section{Introduction}

The $\Lambda(1405)$ is a negative parity excited baryon with strangeness 
$S=-1$. It is well known that simple constituent quark models have a 
difficulty in describing the light mass of the $\Lambda(1405)$, in spite of 
the successful description of other excited baryons. On the other hand, the 
$\Lambda(1405)$ can be reasonably described by the coupled-channel framework
of meson-baryon scattering. These observations have caused a long-standing 
discussion about its structure. A recent interest on this particle is related
to the $\bar{K} N$ interaction below the threshold. It should be noted that 
the subthreshold $\bar{K}N$ amplitude cannot be directly observed in 
experiments since it is kinematically forbidden. The only way to access this 
region is given by the spectrum of the coupled-channel $\pi \Sigma$, which is
largely dominated by the $\Lambda(1405)$ resonance. Therefore the property of
the $\Lambda(1405)$ is related to the $\bar{K} N$ interaction below 
threshold, which is an essential building block for the study of kaonic 
nuclei. Here we report on recent development of chiral SU(3) dynamics about 
these issues: the structure of the $\Lambda(1405)$ and its consequence in 
$\bar{K}N$ phenomenology.

\section{Nonperturbative chiral dynamics}

We utilize the chiral coupled-channel approach, which describes hadron 
scatterings and resonances, incorporating two important 
principles: (i) the interaction follows the low energy theorem of chiral 
symmetry,\cite{WT} and (ii) the amplitude should be constrained by the 
unitarity condition in coupled channels.\cite{Dalitz:1967fp} In practice, we 
determine the low energy interaction by the leading order term of chiral 
perturbation theory:
\begin{equation}
    V_{ij}
    \sim -\frac{C_{ij}}{4f^2}(\omega_i+\omega_j) , 
    \label{eq:WT}
\end{equation}
where $C_{ij}$ are the coupling strengths, $f$ is the meson decay constant, 
and $\omega_i$ is the energy of the meson in channel $i$. The coupled-channel
scattering amplitude $T_{ij}$ is given by solving the Bethe-Salpeter (BS) 
equation
\begin{equation}
    T_{ij}
    =V_{ij}+V_{il}G_lT_{lj} ,
    \label{eq:BS}
\end{equation}
where $G_i$ is the loop function. Applied to the $S=-1$ meson-baryon 
scattering system,\cite{ChU} the resulting amplitude well reproduces the 
experimental data such as total cross sections of $K^-p$ scattering, 
threshold branching ratios, and the invariant mass spectrum in $\pi \Sigma$ 
channel. The framework has been successfully applied also to the $S=0$ sector
with nucleon resonances, to mesonic sectors, and to systems including heavy 
quarks in the target hadrons. These successes in variety of systems indicate 
the importance of the above two principles (i) and (ii) when constructing the
hadron scattering amplitude.\cite{Exotic}

\section{Structure of the $\Lambda(1405)$ resonance}

We first analyze the structure by paying attention to the renormalization 
procedure. Based on a general ground of the scattering theory, origin of 
resonances can be classified into two categories: dynamical state and 
Castillejo-Dalitz-Dyson (CDD) pole contribution.\cite{Castillejo:1956ed} The 
dynamical state is the resonance generated by the two-body interaction, which
is regarded as a two-body molecule state. The CDD pole is considered to be an
elementary or independent particle, which has the origin outside the model 
space of the scattering. In the present case of the $\Lambda(1405)$, the CDD 
pole contribution would come, for instance, from three-quark state. The 
previous studies of the chiral coupled-channel approach have shown that CDD 
pole contribution is introduced in several ways to the interaction kernel 
$V_{ij}$ in Eq.~\eqref{eq:BS}. In a recent work,\cite{Hyodo:2008xr} it is 
pointed out that the CDD pole contribution can also exist in the loop 
function $G_i$ through the renormalization procedure. It has been explicitly 
shown that a certain choice of the cutoff parameter introduces the pole term 
contribution in $V_{ij}$, even if the CDD pole contribution is not included 
in the beginning. Analyzing the two examples, $N(1535)$ resonance in $\pi N$ 
scattering and $\Lambda(1405)$ resonance in $\bar{K} N$ scattering, it is 
found that a substantial CDD pole contribution is required for the $N(1535)$ 
on top of the dynamical component, while the $\Lambda(1405)$ is largely 
dominated by the dynamical component.

Next we consider the scaling with respect to the number of colors $N_c$, 
which is a powerful tool to investigate the quark structure in hadron 
effective theory. The key issue is that, in QCD, the $N_c$ dependences of the
hadronic quantities are known from the general argument.\cite{Nc} Therefore, 
introducing the $N_c$ dependence into the framework and analyzing the 
properties of the resonance with respect to $N_c$, we can extract the 
information of quark structure of the resonance.\cite{Pelaez:2003dy} In 
contrast to the mesonic sector, baryonic sector contains the nontrivial $N_c$
dependence in the leading order WT interaction.\cite{Exotic} Introducing all 
the $N_c$ dependences in the model, the scattering amplitude is calculated as
a function of $N_c$. The general argument tells us that if the excited baryon
is a three-quark state, the mass is proportional to $N_c$ and the width 
should be a constant [$\Gamma\sim \mathcal{O}(1)$]. The result shows that the
width of the $\Lambda(1405)$ resonance changes when the $N_c$ is 
increased.\cite{NcLambda} This is a clear indication of the non-$qqq$ 
structure of the $\Lambda(1405)$ resonance.

Finally we study electromagnetic properties of the $\Lambda(1405)$ resonance 
by introducing an external photon field.\cite{Sekihara:2008qk} Because the 
resonance is expressed by the bubble sum of the meson-baryon loops, the 
photon field is attached to the constituent mesons and baryons. In order to 
extract the information of size, we evaluate the electric mean squared radius
by neglecting the decay channels. The result turns out to be about 2 fm$^2$ 
with negative sign. This means that the electromagnetic size of this 
resonance is much larger than the ground state nucleon, which is mainly 
composed of three-quark state. Hence the result is consistent with the 
picture of the $\Lambda(1405)$ that the $K^-$ is widely spread around the 
proton.

To summarize the above results, we list the possible quark and hadronic 
components of the $\Lambda(1405)$ in Table~1. For an ordinary baryon, the 
main component would be the three-quark state ($qqq$). The existence of the 
meson-baryon component ($MB$) in baryonic system is also known, for instance,
by the pion cloud in the ground state nucleon. There could be further 
contribution from other components on top of $qqq$ and $MB$. Note that we 
distinguish the meson-baryon state $MB$ from five-quark state $qqqq\bar{q}$; 
the former is a combination of two color singlet hadrons, while the latter is
defined as a totally color singlet five-quark state which has no overlap with
the $MB$ state.\cite{Hyodo:2008xr}

The analysis of renormalization procedure\cite{Hyodo:2008xr} implies that the
$\Lambda(1405)$ resonance is dominated by the $MB$ dynamical component. The
result of the $N_c$ scaling\cite{NcLambda} indicates that the three-quark 
component of the $\Lambda(1405)$ is small, although we cannot specify the 
explicit origin of the resonance by itself. The study of the electromagnetic 
properties\cite{Sekihara:2008qk} shows that the size of the $\Lambda(1405)$ 
is larger than the ordinary baryons. It is remarkable that all three 
different analyses are consistent with the dominance of meson-baryon molecule
structure of the $\Lambda(1405)$.

\begin{table}[pt]
\tbl{Schematic classification of the results of the analyses for the 
structure of the $\Lambda(1405)$. Components $q$, $M$, and $B$ stand for
quark, meson, and baryon, respectively.}
{\begin{tabular}{@{}lcccc@{}} \toprule
Components & $qqq$ & $MB$ & other components ($qqqq\bar{q}$, 
$MMB$, \ldots) \\ \colrule
Ref.~\refcite{Hyodo:2008xr} (CDD pole) &  & likely &  \\
Refs.~\refcite{NcLambda} ($N_c$ scaling) & not likely & &  \\
Ref.~\refcite{Sekihara:2008qk} (charge radius) & not likely & &  \\  \botrule
\end{tabular}
}
\end{table}

\section{Effective $\bar{K}N$ interaction and kaonic nuclei}

The possible antikaon binding in nucleus was discussed in Refs.~\refcite{AY} 
using phenomenological $\bar{K}N$ interaction. It was argued that the strong 
$\bar{K}N$ attraction would cause many interesting phenomena in kaonic 
nuclei. Stimulated by recent experimental searches, this topic is now lively 
discussed. Since we have a theoretical framework of chiral dynamics, which 
reproduces the experimental data of $\bar{K} N$ scattering quite well, it is 
natural to ask what chiral dynamics tells us about the kaonic nuclei.

For this purpose, an effective single-channel $\bar{K}N$ potential is derived
in chiral dynamics,\cite{Hyodo:2007jq} which enables a standard variational 
calculation of the few-body kaonic nuclei. The strategy is as follows: (a) we
transform the original coupled-channel framework into the single channel 
problem, and (b) we construct an equivalent potential to be used in the 
Schr\"odinger equation of single channel. In step (a), the effect of the 
$\pi \Sigma$ channel is included in the single-channel effective $\bar{K}N$ 
interaction within an exact transformation. Step (b) requires a local 
approximation for the potential in coordinate space, but we impose the 
constraint that the potential should reproduce the scattering amplitude of 
chiral coupled-channel approach. 

In this way we construct an effective $\bar{K}N$ potential, which 
incorporates the dynamics of $\pi \Sigma$ and reproduces the scattering 
amplitude in chiral dynamics. The strength of the attraction turns out to be 
about a half of the phenomenological potential of Refs.~\refcite{AY}. Indeed,
the variational calculation of $K^-pp$ system with this potential shows a 
small binding energy of about 20 MeV.\cite{Dote} Let us consider the reason 
for the weaker interaction.

Actually the consequence of the weaker attraction is seen in the two-body 
scattering amplitude.\cite{Hyodo:2007jq} The resonance position in the
$\bar{K}N$ amplitude appears at around 1420 MeV, not at the nominal position 
of 1405 MeV. Since the binding energy measured from the $\bar{K}N$ threshold 
is reduced to 15 MeV, the effective $\bar{K}N$ interaction is less 
attractive. The shift of the resonance energy is the key to determine the 
strength of the attraction. Note that the $\pi\Sigma$ amplitude, which 
corresponds to the observed spectrum, shows the resonance structure at around
1405 MeV.

The difference between the resonance positions in the $\bar{K}N$ and
$\pi \Sigma$ channels was discussed in Ref.~\refcite{Jido:2003cb}. It turns 
out that there are two poles for this resonance, and the poles couple to the 
$\bar{K}N$ and $\pi \Sigma$ states with different weights, so that the 
spectrum differs each other. In Ref.~\refcite{Hyodo:2007jq}, it is pointed 
out that the diagonal couplings in Eq.~\eqref{eq:WT} are attractive enough to
generate singularities of scattering amplitude in both $\bar{K}N$ and 
$\pi\Sigma$ channels ($C_{\bar{K}N}=3$ and $C_{\pi\Sigma}=4$). This is in 
contrast to the phenomenological potential,\cite{AY} where the diagonal 
$\pi\Sigma$ interaction is set to be zero. The strong $\pi\Sigma$ attraction 
in chiral dynamics eventually reduce the strength in the equivalent local 
$\bar{K}N$ potential. The coupling strengths in Eq.~\eqref{eq:WT} are 
strictly governed by the flavor SU(3) symmetry and this feature is also 
shared with the traditional coupled-channel approach by 
Dalitz.\cite{Dalitz:1967fp}

From a schematic viewpoint, the phenomenological potential\cite{AY} describes
the $\Lambda(1405)$ as a Feshbach resonance: quasibound $\bar{K}N$ state 
embedded in the $\pi\Sigma$ continuum. In the chiral scheme, the driving 
force to generate the $\Lambda(1405)$ is the attraction in the $\bar{K}N$ 
channel, while the $\pi\Sigma$ system is also strongly and attractively 
interacting. As a consequence, effective $\bar{K}N$ interaction is less 
attractive than the phenomenological one, in order to achieve the same 
spectrum of the $\Lambda(1405)$.

\section{Conclusions}

We have reviewed some recent development in chiral SU(3) dynamics related to 
the $S=-1$ meson-baryon scattering. Within the present framework, we find 
that the structure of the $\Lambda(1405)$ is dominated by the meson-baryon 
molecule component. The effective $\bar{K}N$ interaction turns out to be less
attractive than the phenomenologically constructed one, because of the strong
$\pi\Sigma$ dynamics constrained by chiral SU(3) symmetry.

\section*{Acknowledgements}

The author is grateful to Akinobu Dot\'e, Atsushi Hosaka, Daisuke Jido, Luis 
Roca, Takayasu Sekihara, and Wolfram Weise for fruitful collaborations. He 
thanks the Japan Society for the Promotion of Science (JSPS) for financial 
support. This work is supported in part by the Grant for Scientific Research 
(No.\ 19853500) from the Ministry of Education, Culture, Sports, Science and 
Technology (MEXT) of Japan. This research is part of the Yukawa International
Program for Quark-Hadron Science. 
    
\appendix

\end{document}